\documentclass[12pt]{article}
\usepackage{epsfig}
\newcommand{\bi}{b_i}

\newcommand{\ati}{\alpha_i}
\newcommand{\be}{\begin{equation}}
\newcommand{\ee}{\end{equation}}
\newcommand{\bea}{\begin{eqnarray}}
\newcommand{\eea}{\end{eqnarray}}
\newcommand{\ii}{{_i}}
\newcommand{\jj}{{_j}}
\newcommand{\brho}{{\bar \rho}}
\newcommand{\bp}{{\bar p}}
\newcommand{\bP}{{\bar P}}

\newcommand{\Lhat}{\widehat{\mathcal{L}}}
\newcommand{\Gn}{G_{N+n+1}} 
\newcommand{\mpl}{M_P}
\newcommand{\rt}{\sqrt{3}}
\newcommand{\mphi}{\mu_\Phi^2}
\newcommand{\mpsi}{\mu_\Psi^2}
\title{Classical Stabilization of Homogeneous Extra Dimensions}
\author{Sean M. Carroll,\thanks{\tt carroll@theory.uchicago.edu} \ James
Geddes,\thanks{\tt jm-geddes@uchicago.edu} 
\ Mark B. Hoffman,\thanks{\tt mb-hoffman@uchicago.edu} \ and Robert
M. Wald\thanks{\tt rmwa@midway.uchicago.edu} \\ \\
\it Enrico Fermi Institute and Department of Physics, \\
\it University of Chicago \\
\it 5640 S.~Ellis Avenue, Chicago, IL~60637, USA \\ \\
hep-th/0110149. EFI-2001-38}
\begin{document}
\baselineskip 20pt
\maketitle

\begin{abstract}

If spacetime possesses extra dimensions of size and curvature radii
much larger than the Planck or string scales, the dynamics of these
extra dimensions should be governed by classical general relativity.
We argue that in general relativity, it is nontrivial to obtain
solutions where the extra dimensions are static and are dynamically
stable to small perturbations. We also illustrate that intuition on
equilibrium and stability built up from non-gravitational physics can
be highly misleading. For all static, homogeneous solutions satisfying
the null energy condition, we show that the Ricci curvature of space
must be nonnegative in all directions.  Much of our analysis focuses
on a class of spacetime models where space consists of a product of
homogeneous and isotropic geometries. A dimensional reduction of these
models is performed, and their stability to perturbations that
preserve the spatial symmetries is analyzed.  We conclude that the
only physically realistic examples of classically stabilized large
extra dimensions are those in which the extra-dimensional manifold is
positively curved.

\end{abstract}

\vfill\eject

\section{Introduction}

The idea that spacetime may have more than 4 dimensions has had a long
and distinguished history. In the context of string theory or other
quantum theories of gravity, the extra dimensions typically would
be expected to be of a size set by the string or Planck length. In
such cases, one would not expect a classical description of the extra
dimensions to be valid. However, in recent years, the possibility has
been raised that ``large'' extra dimensions may be present, and a
number of such explicit models have been proposed (\cite{add};
see \cite{review} for a review, and \cite{exp} for recent experimental
limits). It seems reasonable
to expect that classical general relativity should be adequate to
analyze the equilibrium and stability of these models. 

In general relativity, it is nontrivial to obtain models with static
extra dimensions, and it is even more nontrivial to stabilize these
extra dimensions with respect to small perturbations. Furthermore,
intuition developed from non-gravitational physics with regard to
equilibrium and stability can be highly misleading. Indeed, the
difficulties in obtaining stable compact dimensions and the
counter-intuitive behavior of many models are well illustrated by a
simple example.

Consider a static, flat, $n$-torus universe. (In this example, the
$n$-torus will be taken to be all of space, but a similar example
could be given where the $n$-torus comprises only the extra
dimensions.) This spacetime is, of course, a static, equilibrium
solution of the vacuum Einstein equation. Is this solution stable? The
following argument suggests that it should not be: There is no scale
that sets the values of the ``moduli'' of the torus. In particular,
there is an equilibrium solution for all possible values of the radii
of the torus. Therefore, at best, one would expect the torus to be
neutrally (un)stable: If one of the dimensions of the torus is
perturbed toward collapse, there should be nothing to prevent that
dimension from collapsing to zero size in a finite time. It is not
difficult to prove that this expectation is correct: There exist exact
solutions of the vacuum Einstein's equation corresponding to a linear
decrease (or growth) of the size of one of the dimensions of the torus
with time,\footnote{These exact solutions with linear expansion or
contraction of one of the radii of the torus are actually flat. They
can be obtained by starting with a wedge of $(1+1)$ dimensional
Minkowski spacetime where the action of Lorentz boosts is spacelike
(the ``Milne universe'') and making a periodic identification under
Lorentz boosts.} so an arbitrarily small perturbation will cause the
torus to collapse within a finite time.

Can the above $n$-torus universe be stabilized via the addition of
suitable matter fields? One might expect that it would be
straightforward to do so by simply adding some matter---such as iron
struts---whose total energy was minimized at a particular chosen set
of values of the radii of the torus. This matter would then set a
suitable equilibrium scale for the torus, and one might expect that
the $n$-torus universe at the energy minimizing radii would be a
stable, equilibrium configuration.  (In the somewhat different
context of a single extra dimension
with orbifold boundary conditions, this is essentially the
idea of the Goldberger-Wise stabilization
mechanism~\cite{goldbergerwise}.)  However, it is easy to see that this
mechanism does not work in this case. In fact, we will see from
eq.~(\ref{intcons3}) below that if one adds matter satisfying the
strong energy condition with strict inequality holding at at least one
point, then no static solutions whatsoever can exist, so all solutions
must be expanding or contracting in at least one spatial
dimension. Furthermore, if we consider a contracting solution with the
same rate of contraction in all directions (thereby corresponding to a
flat FRW solution with periodic identifications), then it also can be
seen that the presence of a positive pressure enhances the rate of
collapse in the sense that, with the same initial conditions, a
universe filled with matter possessing positive pressure will collapse
to zero size faster than a universe filled with dust. Thus, rather
than stabilizing the torus, adding ``iron struts'' or other matter
with positive pressure actually enhances the tendency to collapse.

The difficulties in achieving stable equilibrium of compact dimensions
are not special to the above $n$-torus example. Indeed, a simple
constraint on the existence of stationary solutions can be obtained as
follows. Consider a stationary spacetime, with timelike Killing field
$t^a$, which is spatially compact or compactifiable
(i.e., possessing appropriate spatial
symmetries so that the noncompact directions could be compactified via
identifications). This encompasses many models considered in the
literature, including all of the ones we shall consider below in
this paper.  Then $t^a$ satisfies (see, e.g., Appendix~C of
\cite{waldbook})
\be\label{Killing}
\nabla^a \nabla_a t_b  = - R_{bc} t^c
\ee
Since $\nabla_a t_b = \nabla_{[a} t_{b]}$, this equation may be
rewritten in differential forms notation as
\be\label{Killing2}
d*dt = -2 * (R \cdot t)
\ee
where $(R \cdot t)_a \equiv R_{ab} t^b$. Integrating this equation
over a compact (or compactified) spacelike slice $\Sigma$, we obtain
\be\label{intcons}
\int_\Sigma * (R \cdot t) = 0
\ee
i.e.,
\be\label{intcons2}
\int_\Sigma R_{ab} n^a t^b d\Sigma = 0
\ee
where $n^a$ denotes the unit normal to $\Sigma$, and $d\Sigma$
denotes the volume element on $\Sigma$ induced from the spatial metric
on $\Sigma$. In particular, if the spacetime is static (rather than
merely stationary) and we choose $\Sigma$ to be orthogonal to $t^a$,
then we obtain
\be\label{intcons3}
\int_\Sigma R_{ab} n^a n^b V d\Sigma = 0 ,
\ee
where $V = (- t^a t_a)^{1/2}$.

Eq.~(\ref{intcons2}) and/or eq.~(\ref{intcons3}) provide a
significant integral constraint on the existence of stationary
solutions with compact or compactifiable spacelike slices. 
The strong energy condition is the requirement that
$(T_{ab} - \frac{1}{2}Tg_{ab})v^a v^b \geq 0$ for all timelike
$v^a$. It is violated by a positive cosmological constant and by
matter fields whose stress-energy is dominated by (positive) potential
energy, but is satisfied by most other forms of matter usually
considered. In the presence of Einstein's equation, the strong energy
condition is equivalent to the condition that $R_{ab} v^a v^b \geq 0$
for all timelike $v^a$.  Eqs.~(\ref{intcons2}) and (\ref{intcons3})
indicate that it will be difficult to produce stationary solutions to
Einstein's equation with matter without violating the strong energy
condition. In particular, as already mentioned above, in the static
case, if there is a point where $R_{ab} n^a n^b > 0$, then there also
must be a point where $R_{ab} n^a n^b < 0$, thereby violating the
strong energy condition.  

Similar conclusions concerning the difficulties with the existence of
static solutions satisfying the strong energy condition can be drawn
by appealing to the singularity theorems. In particular, Theorem~9.5.4 of
\cite{waldbook} asserts that if a spacetime possesses a compact
spatial slice, if matter satisfies the strong energy condition, and
if, in addition, there are no closed timelike curves and the timelike
and null generic conditions hold, then the spacetime must be singular
in the sense of timelike or null geodesic
incompleteness. Consequently, all stationary globally hyperbolic
spacetimes possessing a compact or compactifiable Cauchy
surface---such as the static $n$-torus discussed above---must violate
the generic condition. By the above theorem, any perturbation of such
a spacetime that satisfies the generic condition must produce a
singularity.

The above arguments against the existence of stable, static solutions
can be evaded by allowing for the presence of a positive cosmological
constant or other forms of matter violating the strong energy
condition. Even so, it is still far from obvious how to make stable,
static models. Indeed, the Einstein static universe provides a good
illustration of the kind of difficulties one can encounter: Although a
static solution can be obtained by ``balancing'' a positive
cosmological with an ordinary form of matter, such an
equilibrium is unstable.

The main purpose of this paper is to investigate the constraints
placed by classical general relativity on the existence and stability
of static solutions to Einstein's equation in the case where ``space''
is taken to be a product of an arbitrary number of homogeneous and
isotropic geometries of arbitrary curvature---in other words,
conventional Kaluza-Klein theory \cite{kk}.  In sections 2 and 3, we
restrict attention to the case of a product of two such
geometries---one of which can be taken to be ordinary, macroscopic
space, and the other of which can be taken to be the ``large''
(compared with the Planck scale) extra dimensions. In section 2, we
show that if matter satisfies the null energy condition ($T_{ab} l^a
l^b \geq 0$ for all null $l^a$), then existence of a static solution
requires either that the compact space be positively curved, or that
the spacetime be locally Minkowski and the stress-energy tensor vanish
identically.  Thus, provided only that the null energy condition
holds, the extra dimensions in the static models with the symmetries
we consider cannot have negative curvature, and they cannot even be
flat unless the universe is completely empty. More generally, we show
that for any static, homogeneous solution with matter satisfying the
null energy condition, the Ricci curvature of space must always be
nonnegative.

In section 3, we consider a particular model that has previously been
considered in the literature~\cite{hpcs,freundrubin,rss,gz1,sundrum},
in which spherical extra dimensions are stabilized via a balance
between vacuum energy and gauge fields (and the curvature of the
internal space).  We perform a dimensional reduction of it, thereby
reducing it to an ordinary Robertson-Walker model with a
self-interacting scalar field. The static solution is unstable in the
sense that an arbitrarily small perturbation of it will result in a
(``big bang'') singularity. However, the perturbed model appears to be
cosmologically acceptable.

Section 4 is devoted to exploring the relationship between energy
minimization versus equilibrium and stability in general relativity.
In non-gravitational physics, a configuration which minimizes the
total matter energy corresponds to a stable equilibrium configuration,
and stability arguments are typically made on the basis of energy
minimization. However, such arguments are, at best, highly suspect in
general relativity. In asymptotically flat spacetimes---where total
energy is well defined---configurations of minimum total energy should
correspond to stable equilibrium configurations~\cite{sudarskywald},
but the total energy of the spacetime is not equal to the integrated
energy of matter in the spacetime, so minimization of the total {\it
matter} energy may have little to do with the conditions for stable
equilibrium. The situation is considerably worse in compact or
compactifiable spaces, where no nontrivial notion of total energy
exists. To investigate the relationship between minimization of total
matter energy and stable equilibrium, we consider a Robertson-Walker
spacetime filled with a fluid, and ask the following two questions:
(i) What conditions on the pressure of the fluid must be satisfied in
order that the total energy of the fluid be minimized?  (ii) What
conditions on the pressure of the fluid must be satisfied in order to
have a stable, static solution to Einstein's equation?  We shall see
in section 4 that, in general, the conditions for equilibrium in
general relativity bear no relationship with the conditions for energy
extremization. Interestingly, however, in $2+1$ dimensions, it turns
out that the conditions for equilibrium in general relativity actually
coincide with the conditions for energy extremization. Remarkably, in
this case we find that the conditions for {\em stable} equilibrium in
general relativity are precisely that the matter energy be a local
{\em maximum}---exactly the opposite of what would be expected in
non-gravitational physics.

Finally, in section 5, we generalize the results of sections 2 and 3
to the case where the spatial geometry is a product of an arbitrary
number of homogeneous and isotropic geometries. We show that, again,
if matter satisfies the null energy condition, the existence
of a static solution requires nonnegative curvature of {\it each} of 
the geometries; it is not possible to compensate for some negative
curvature by the addition of positive curvature in a different
factor.  Following \cite{gz1,gz2} we also perform a
dimensional reduction in this case, re-expressing the model as
an ordinary Robertson-Walker cosmology with a collection of
self-interacting scalar fields.  Considering a specific example
with four extra dimensions in the form of a product of
two two-spheres \cite{gz1}, we investigate the cosmological
acceptability of such compactifications.

Our analysis leaves numerous questions unanswered.  We do not, for
example, consider off-diagonal perturbations to our product spacetimes
or perturbations that do not respect the spatial symmetries of the
extra dimensions.  Also, models where the standard-model fields are
confined to a ``brane'' with a thickness much smaller than the size of
the extra dimensions are incompatible with our assumption of perfect
homogeneity in the extra dimensions.  (In particular, Randall-Sundrum
spacetimes with a single finite extra dimension
\cite{rs1} rely on the tension of the branes on the boundary, and
are therefore not addressed by our analysis.)  While a natural first step
is to assume that the energy localized on the brane is smaller than
that in the ``bulk'' and can hopefully be neglected, it is important
to subsequently check that the introduction of a nonzero brane tension
does not destabilize the solution.  (Some progress in including 
the effects of nonzero brane tension has been made
by considering metrics generated by global topological defects
\cite{defects}.)

\section{Static, homogeneous solutions in general relativity}

In this section we examine static homogeneous spacetimes, without
regard to questions of stability (considered in later sections).
We consider spacetimes of the form
${\bf R}\times {\cal M}_1 \times {\cal M}_2$, where ${\bf R}$ is
the timelike direction and ${\cal M}_1$ and ${\cal M}_2$
are maximally symmetric manifolds of dimensionality $N$ and $n$,
respectively.  We can think of ${\cal M}_1$ as representing the 
``large'' spatial dimensions and ${\cal M}_2$ as representing
the ``extra'' ones, although they will be treated on an identical
footing.  The metric can be written
\be
  ds^2 = -dt^2 + g_{IJ}dx^I dx^J + 
  \gamma_{ij}dy^i dy^j ,
\ee
where $g_{IJ}(x^K)$ and $\gamma_{ij}(y^k)$ are metrics on manifolds of
constant curvature (either positive, negative, or zero).
Note in particular that $g_{IJ}(x^K)$ and $\gamma_{ij}(y^k)$ are
independent of time.
Indices $I,J$ run from $1$ to $N$, while $i,j$ run from $1$ to $n$.
We define curvature parameters by
\be
  K = \frac{R[g_{IJ}]}{N(N-1)} ,\qquad
  k = \frac{R[\gamma_{ij}]}{n(n-1)} ,
  \label{kdef}
\ee
where $R$ is the Ricci scalar constructed from the appropriate
spatial metric.

For a stress-energy source we take a perfect fluid defined
by an energy density
$\rho$ and pressure in the large and small dimensions, $P$
and $p$:
\bea
  T_{00} & = & \rho \\
  T_{IJ} & = & P g_{IJ} \\
  T_{ij} & = & p \gamma_{ij} .
\eea
There are three independent nontrivial components of
Einstein's equation,
corresponding to the $00$, $IJ$ and $ij$ components:
\bea
  \frac{1}{2}N(N-1)K + \frac{1}{2}n(n-1)k & = & 8\pi G \brho 
  \label{static1}
  \\
  \frac{1}{2}(N-1)(N-2)K + \frac{1}{2}n(n-1)k & = & -8\pi G \bP 
  \label{static2}
  \\
   \frac{1}{2}N(N-1)K + \frac{1}{2}(n-1)(n-2)k & = & -8\pi G \bp , 
  \label{static3}
\eea
where $G$ is Newton's constant (in $N+n+1$ dimensions).  In
(\ref{static1}-\ref{static3}) we have put overbars on the energy density
and pressures to emphasize that these relations hold at a
static solution.

We next impose the null
energy condition\footnote{The NEC is in fact weaker than
the ``weak'' energy condition, that $T_{ab}v^a v^b \geq 0$
for all timelike vectors $v^a$.  By continuity the weak energy
condition implies the NEC, but not vice-versa.  In particular, the
NEC permits a negative energy density if there is an equal and
opposite pressure, and thus allows for a cosmological constant of
either sign, while the weak energy condition prohibits any negative
energy density, and thus allows only a nonnegative cosmological
constant.} (NEC), that $T_{ab}l^al^b \geq 0$
for all null vectors $l^a$.  In the context of our model, the NEC is
equivalent to the requirements
\bea
  \rho + P &\geq & 0 ,
  \label{nec1} \\
  \rho + p &\geq & 0 .
  \label{nec2}
\eea
Taking appropriate linear combinations of
(\ref{static1}-\ref{static3}) yields
\bea
  8\pi G(\brho + \bP) &=& (N-1)K ,
  \label{static4}\\
  8\pi G(\brho + \bp) &=& (n-1)k  .
  \label{static5}
\eea
From comparing (\ref{static4}-\ref{static5}) with
(\ref{nec1}-\ref{nec2}), we see immediately that the curvatures
$K$, $k$ must be either zero or positive.  (The curvature cannot
be made negative by choosing a one-dimensional manifold,
as a one-dimensional space must have zero curvature.)  
Thus, in a static spacetime obeying the NEC, with spatial 
sections described by a product of two symmetric spaces, neither
spatial factor can be negatively curved.  In particular,
the NEC must be violated in order to consider stabilization
of compact hyperbolic extra dimensions, which have recently
been investigated \cite{chms}.

Several possibilities remain open:  both spaces may be flat,
one may be flat and the other positively curved, or both may
be positively curved.  If both spaces are flat ($K=k=0$),
we see from (\ref{static1}-\ref{static3}) that all of the
components of the stress-energy tensor must vanish,
$\brho = \bP = \bp = 0$.  This solution is (locally) empty
Minkowski space.

The nonempty static solutions obeying the NEC are thus those
in which at least one space is positively curved, and neither
is negatively curved.  This implies that toroidal compactification
($k=0$) is only possible if the large dimensions are positively
curved ($K>0$).  In fact, however, this case is not physically
realistic, for two reasons.  First, in the real universe the three
observed spatial dimensions are flat to an excellent approximation.
(Evidence for spatial flatness has been provided by measurements
of the anisotropy spectrum of the cosmic microwave background
\cite{cmb}; for our present purposes, however, all we need to
know is that radius of curvature of the large dimensions is
significantly larger than any relevant microphysical scale.)
Second, when (in the next section) we consider stability of
a model with both flat and positively-curved dimensions, we will
find that small perturbations induce oscillations in the positively
curved space and a big-bang singularity in the flat dimensions.
Hence, a toroidal set of extra dimensions would be unstable
to collapsing to zero volume (or expanding to infinite volume).
We therefore conclude that the only physically relevant examples
of large extra dimensions obeying the NEC are those with 
positive curvature.

The above result on the non-negativity of spatial curvature for static
solutions can be significantly generalized as follows. Consider any
static, spatially homogeneous spacetime, with static Killing field
$t^a$. Now, any static Killing field satisfies
\be\label{staticKilling}
\nabla_a t_b  = - 2 V^{-1} t_{[a} \nabla_{b]} V 
\ee
where $V = (- t^a t_a)^{1/2}$. However, for a spatially homogeneous
spacetime, we have $\nabla_a V = 0$, so $t^a$ is covariantly
constant. It then follows from eq.~(\ref{Killing}) that $R_{ab} t^b =
0$, where $R_{ab}$ denotes the Ricci curvature tensor of the spacetime
metric. Let $s^a$ be any unit ``spatial'' vector (i.e., $t^a s_a = 0$),
and let $l^a = V^{-1}t^a + s^a$. Then $R_{ab} l^a l^b = R_{ab} s^a
s^b$. However, since $l^a$ is null, by Einstein's equation, we have
\be\label{nullEinstein}
 R_{ab} l^a l^b = 8\pi G(T_{ab} - \frac{1}{2} g_{ab} T) l^a l^b = 8\pi G
 T_{ab} l^a l^b
\ee
Consequently, the NEC together with Einstein's equation implies
$R_{ab} s^a s^b \geq 0$ for all spatial vectors $s^a$, with equality
holding if and only if $T_{ab} l^a l^b = 0$. Furthermore, since $t^a$
is covariantly constant, we have $R_{ab} s^a s^b = \tilde{R}_{ab} s^a
s^b$, where $\tilde{R}_{ab}$ denotes the Ricci curvature of space.
Thus, for any static, spatially homogeneous spacetime satisfying the
null energy condition, the Ricci curvature of space must always be
nonnegative.

\section{An example: stabilizing $S^2$ with gauge fields}
\label{example}

Having established the criteria for static solutions with a single
extra-dimensional space, we now will examine in detail a specific
example to understand the sense in which it can be ``stable''.  Our
example will be a spacetime with four macroscopic spacetime dimensions
and two positively curved extra dimensions, stabilized by
balancing a cosmological
constant and a gauge field wrapped around the extra dimensions. 
(This mechanism was recently examined by Sundrum \cite{sundrum};
a nonabelian version was originally investigated by Horvath,
Palla, Cremmer and Scherk \cite{hpcs}, 
which was generalized to larger numbers of extra
dimensions by Freund and Rubin \cite{freundrubin}, and stability
was discussed by Randjbar-Daemi, Salam and Strathdee \cite{rss}.
A thorough discussion, including multiple compact spaces, is
given by Guenther and Zhuk \cite{gz1}.)  We
consider the conventional dimensional reduction of this theory,
rewriting the six-dimensional Lagrangian as a four-dimensional
Lagrangian plus a scalar field (the radion) representing the size of
the extra dimensions.  The effective potential for the radion will be
found to have two minima: one at a finite radius, and one at infinite
radius.  This kind of analysis is by no means original (see for example
\cite{overduinwesson} and references therein), but we will
go through it carefully for pedagogical purposes.

Let $G_{ab}$ be the metric for the full spacetime
with coordinates $X^a$, and denote by $\Gn$
Newton's constant in the full spacetime. We consider
$N$ large spatial dimensions and $n$ extra dimensions, 
so that indices $a$, $b$ run from $0$ to
$N+n$; we will specialize to the case $N = 3$, $n = 2$ after
performing the dimensional reduction. We consider metrics of the form
\be
  ds^2 = G_{ab}dX^adX^b = g_{AB}(x)dx^Adx^B + b^2(x)\gamma_{ij}(y)dy^i dy^j,
  \label{metricansatz}
\ee
where the $x^A$ are coordinates in the $(N+1)$-dimensional spacetime and
the $y^i$ are coordinates on the extra-dimensional manifold, again
taken to be a maximally symmetric space with metric $\gamma_{ij}$.
However, for the dimensional reduction analysis given below, it will {\em
not} be assumed that the metric $g_{AB}$ of the large dimensions has
any symmetries.
The action is the $(N+n+1)$-dimensional Hilbert action, plus a matter
term we leave unspecified for now:
\begin{equation}\label{eq:HilbertAction} 
  S = \int d^{N+n+1}X \sqrt{-G}\left\{\frac{1}{16\pi\Gn} R[G_{ab}]
   +\Lhat_M\right\},
\end{equation}
where $\sqrt{-G}$ is minus the square root of the determinant of 
$G_{ab}$, $R[G_{ab}]$ is the Ricci scalar of $G_{ab}$, and $\Lhat_M$ is
the matter Lagrangian density with the metric determinant factored out.

The first step is to dimensionally reduce the action 
(\ref{eq:HilbertAction}):
we express everything in terms of $g_{AB}$, $\gamma_{IJ}$, and
$b(x)$, and integrate over the extra dimensions
\footnote{As is well known, if one substitutes a metric (or other field)
ansatz into an action, there is a danger that variation of the new action
will not reproduce all of the field equations, since the metric (or other
field) variations are restricted by the ansatz. If the ansatz arises from
symmetry assumptions, then, typically, the missing field equations will be
automatically satisfied by virtue of the assumed symmetry. It is not
difficult to verify that this is the case here, so that the reduced action
that we will obtain below does indeed yield all of Einstein's equations.
On the other hand, if the ansatz arises, e.g., from a gauge choice, then
there is no reason to expect that the new action will reproduce all of the
field equations. An example of this can be found in the last reference
of \cite{add}, where the gauge choice $g_{00} = 1$ is imposed in the
action, and the action consequently fails to yield the $00$-component of
Einstein's equation.}.  From the metric
ansatz (\ref{metricansatz}) we have
\bea
  \sqrt{-G} &=& b^n \sqrt{-g}\sqrt{\gamma}, \\
  R[G_{ab}] &=& R[g_{AB}] + b^{-2} R[\gamma_{ij}] -2n b^{-1}
  g^{AC}\nabla_A\nabla_Cb - n(n-1) b^{-2} g^{AC}(\nabla_A b)
  (\nabla_C b) ,
\eea
where $\nabla_A$ is the covariant derivative associated with the
$(N+1)$-dimensional metric $g_{AB}$.  We denote by $\mathcal{V}$ the
volume of the extra dimensions when $b=1$; it is given by
\be
  \mathcal{V} = \int d^ny \sqrt{\gamma} .
\ee
The ($N+1$)-dimensional Newton's constant $G_{N+1}$ is related to its
higher-dimensional version by
\be
  \frac{1}{16\pi G_{N+1}} = \frac{\mathcal{V}}{16\pi \Gn}.
\ee
We are thus left with
\bea
  S &=& \int d^{N+1}x \sqrt{-g}\Biggl\{\frac{1}{16\pi G_{N+1}} 
  \Bigl[b^n R[g_{AB}] 
  - 2nb^{n-1}g^{AB}\nabla_A\nabla_B b  \nonumber \\
  &&\qquad\qquad\qquad - n(n-1) b^{n-2} g^{AB}(\nabla_Ab)
  (\nabla_Bb)
  + n(n-1)k b^{n-2} \Bigr] + \mathcal{V} b^n \Lhat_M\Biggr\},
  \label{reducedaction}
\eea
where the curvature parameter $k$ of
$\gamma_{ij}$ is given by (\ref{kdef}).

Through a change of variables and a
conformal transformation, 
\bea
  b(x) &=& e^{\beta(x)}, \\
  g_{AB} &=& e^{-2n\beta/(N-1)}\tilde{g}_{AB},
\eea
we can turn this reduced action into that of a scalar field
coupled to gravity in the so-called ``Einstein frame'' (where
the gravitational Lagrangian is simply the Ricci scalar, with
no multiplicative scalar fields).  Some algebra yields
\bea
  S &=& \int d^{N+1}x \sqrt{-\tilde{g}}\Biggl\{\frac{1}{16\pi G_{N+1}} 
  \Bigl[R[\tilde{g}_{AB}]  
  - \frac{n(N+n-1)}{N-1} 
  \tilde{g}^{AB}(\tilde{\nabla}_A\beta)(\tilde{\nabla}_B\beta)
    \nonumber \\
  &&\qquad\qquad\qquad\qquad\qquad
  + n(n-1)k e^{-2\frac{N+n-1}{N-1}\beta} \Bigr] 
	+ \mathcal{V} e^{-2\frac{n}{N-1}\beta} \Lhat_M\Biggr\},
  \label{reducedaction2}
\eea
where we have dropped terms that are total derivatives.  

To turn $\beta$ into a canonically normalized scalar field, we make one
final change of variables, to
\be
  \phi = \sqrt{\frac{n(N+n-1)}{N-1}} M_P \beta ,
\ee
where the reduced Planck mass is $M_P^{-1}= \sqrt{8\pi G_{N+1}}$ (in
units where $\hbar = 1$).  We are then left with
\bea
  S &=& \int d^{N+1}x \sqrt{-\tilde{g}}\Biggl\{\frac{1}{2}M_P^2
  R[\tilde{g}_{AB}] 
  - \frac{1}{2}\tilde{g}^{AB}(\tilde{\nabla}_A\phi)(\tilde{\nabla}_B\phi)
    \nonumber \\
  &&\qquad\qquad\qquad
  + \frac{1}{2}kn(n-1)M_P^2 e^{-2\sqrt{\frac{N+n-1}{n(N-1)}}\phi/M_P}
  + \mathcal{V} e^{-2\sqrt{\frac{n}{(N-1)(N+n-1)}}\phi/M_P} \Lhat_M\Biggr\}.
  \label{reducedaction3}
\eea
Specializing to the case $n=2$ and $N=3$, we see that $\phi$ is a
canonically normalized scalar with a potential
\be
  V(\phi) = -k M_P^2 e^{-2\phi/M_P}
  - \mathcal{V} e^{-\phi/M_P}\Lhat_M(\phi) .
  \label{potential1}
\ee
The stability properties and cosmological evolution of this
model will evidently depend on the choice of matter Lagrangian
$\Lhat_M(\phi)$.

As an attempt at a realistic scenario, we take $\Lhat_M$
to consist of a cosmological constant plus an abelian gauge field 
wrapped around the extra dimensions.  In the six-dimensional
picture, we have an antisymmetric field strength $F_{ab}$ which
we take to be proportional to the volume form of the 
extra-dimensional metric $\gamma_{ij}$; in components we have
\be
  F_{45} = -F_{54} = \sqrt{\gamma} F_0,
\ee
where $F_0$ is a constant and all other components vanish.
In particular, the components of $F_{ab}$ are independent of $b$.  
(It is straightforward to check
that such a field satisfies Maxwell's equations, either directly
or by using conservation of flux.)  We have a 
six-dimensional matter Lagrangian
\be
  \Lhat_M = -\lambda_{6} - \frac{1}{4}G^{ac}G^{bd}F_{ab}F_{cd},
\ee
where $\lambda_6$ is the six-dimensional vacuum energy density.
Upon dimensional reduction this yields
\be
  \mathcal{V}\Lhat_M = -\lambda_4 - f_0^2 e^{-2\phi/M_P},
\ee
where $\lambda_4=\mathcal{V}\lambda_6$ is the 
``bare'' four-dimensional vacuum energy density and 
$f_0^2 = \frac{1}{2}\mathcal{V} F_0^2$ is a (nonnegative) constant.

\begin{figure}
\begin{center}
\epsfig{file=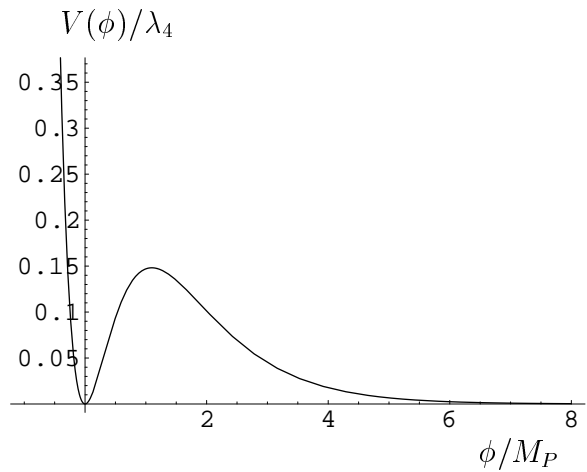}
\end{center}
\caption{\label{fig:effectiveV} The effective potential 
(\ref{potential2}) of the radion
in the model described in the text.}
\end{figure}

The effective potential for $\phi$ in this model is therefore
\be
  V(\phi) = -k M_P^2 e^{-2\phi/M_P}
  + \lambda_4 e^{-\phi/M_P} + f_0^2 e^{-3\phi/M_P}  .
  \label{potential2}
\ee
We would like to choose the parameters $\lambda_4$ and 
$f_0$ such that there is a local minimum of the potential
for which $V=0$ (so that the effective four-dimensional
cosmological constant vanishes and we can have a flat-spacetime
solution).
It is clear from (\ref{potential2}) that choosing $k \leq 0$ 
cannot lead to an isolated minimum at finite $\phi$, in accordance
with the results of the previous section.  
Instead we choose $k > 0$ and
\be
  \lambda_4 = f_0^2 = \frac{1}{2}k M_P^2 .
\ee
There is then a local minimum at $\phi=0$, and the
potential takes the form shown in Figure~\ref{fig:effectiveV};
the mass of the radion is
\be
  m_\phi^2 = \frac{d^2V}{d\phi^2}\Big|_{\phi=0} = k .
  \label{radionmass}
\ee

Note that there is also a minimum as $\phi$ goes to infinity,
corresponding to blowing up the size of the extra dimensions.  This
raises the possibility of a Brustein-Steinhardt
problem~\cite{brusteinsteinhardt}, if cosmological evolution tends to
overshoot the desired minimum at $\phi=0$.  However, this does not
seem like a serious obstacle to the most popular scenarios, which
feature two extra dimensions, a six-dimensional Planck scale of order
1~TeV, and a volume $\mathcal{V}$ of order $(1~{\rm mm})^2 \sim
(10^{-3}~{\rm eV})^{-2}$, so that the four-dimensional Planck scale is
$M_P\sim 10^{18}$~GeV \cite{add}.  From (\ref{radionmass}) the radion
is a light field, with mass set by the length scale of the extra
dimensions.  Meanwhile, the height of the barrier in
Figure~\ref{fig:effectiveV} is of order $V(\phi=M_P)\sim k M_P^2 \sim
1$~TeV$^4$.  Thus, in order for the field to roll coherently over the
barrier, it would have to have an energy density of order the
(six-dimensional) scale of quantum gravity in this model, which is in
a regime in which our classical analysis has no right to be
applicable. However, another potential difficulty with this model
arises from the possibility that perturbations in the early universe
could lead to localized regions in which $\phi$ tunnels over the
barrier, either due to thermal or quantum fluctuations, giving rise to
``bubbles'' in which the extra dimensions expand to infinity.  We do
not analyze this issue here.

Thus, with this model, the effect of the extra dimensions is to 
introduce an additional massive scalar field (the radion)
into the four-dimensional theory.  It is therefore straightforward
to answer questions concerning stability; if we consider a solution
in which the radion is slightly perturbed from its equilibrium
value, it will act as an energy source in the large dimensions,
which will evolve as a Friedmann-Robertson-Walker spacetime.  This
spacetime will have a Big-Bang singularity, but is nevertheless
cosmologically acceptable.  The radion acts as a massive scalar
which contributes to the total energy density of the universe;
so long as its contribution is not too large (or it can decay away),
it will not be cosmologically harmful.

\section{Stabilization vs.\ minimization of energy}

When one is attempting to stabilize the extra dimensions by adding
some form of matter, it might appear natural to expect that the
extra dimensions will be stable when the total matter energy is minimized
(the total matter energy being the integral of the energy density of
matter over the volume). Indeed, this conclusion is rigorously correct
for systems normally considered in non-gravitational physics: the total
energy of matter normally is positive definite (thereby providing an
appropriate norm) and conserved, and only positive energy can be
radiated to infinity. Consequently a small perturbation about the
state of minimum energy cannot drive the system far from this
state. However, this argument fails in general relativity, since there
is no global conservation law for the total energy of
matter\footnote{Inflation provides a dramatic illustration of this
fact.}. The main purpose of this section is to consider a simple
cosmological model illustrating how one can be led seriously astray 
by non-gravitational intuition when determining the stability of
spacetimes in general relativity.

Consider an $(n\!+\!1)$-dimensional spacetime of constant spatial
curvature $k$, so that the metric takes the conventional
Robertson-Walker form
\begin{equation}
  ds^2 = -dt^2 + a(t)^2\gamma_{\alpha\beta}dx^\alpha dx^\beta.  
  \label{rwmetric}
\end{equation}
This space is to be filled with a uniform, perfect fluid of energy density
$\rho$, whose pressure, $p$, is a function of $\rho$. By conservation of
stress-energy, we have
\begin{equation}
p = -\rho - \frac{a}{n}\frac{d\rho}{da},
\end{equation}
This equation can be integrated to express $\rho$ (and, hence, $p$) in
terms of $a$, and, in the following, it will be convenient to view $\rho$
and $p$ as functions of $a$. The quantity $U = \rho a^n$ represents the
``total energy'' (per comoving volume element) of the fluid. Minimizing
$U$ with respect to $a$, we obtain
\begin{equation}
\frac{dU}{da} = -pna^{n-1}.
\end{equation}
Thus for an extremum of $U$ we must have $p=0$. Furthermore, for this
extremum to be a minimum, we require $d^2U/da^2 > 0$, giving
\begin{equation}
\biggl.\frac{d^2U}{da^2}\biggr|_{p=0} = -na^{n-1}\frac{dp}{da} > 0,
\label{enminstab}
\end{equation}
which of course is just the condition that $dp/da$ be negative.
Thus, analysis of the total matter energy in the manner normally
done in non-gravitational physics suggests that space can be in
equilibrium at some size only when the pressure is zero, and stable
only if the pressure increases as the volume shrinks
(``the more the system is squeezed, the harder it pushes back'').

But now consider the analysis of equilibrium and stability in general
relativity, as determined by Einstein's equation.  For the 
Robertson-Walker metric (\ref{rwmetric}) we obtain the Friedmann
equations in $(n\!+\!1)$ dimensions,
\begin{eqnarray}
\frac{1}{2}n(n-1)\Bigl(\frac{\dot{a}}{a}\Bigr)^2 
        &=& 8\pi G\rho-\frac{1}{2}n(n-1)k, \\
  (n-1)\Bigl(\frac{\ddot{a}}{a}\Bigr)
        +\frac{1}{2}(n-1)(n-2)\Bigl(\frac{\dot{a}}{a}\Bigr)^2
  &=& -8\pi Gp -(n-1)\bigl(\frac{n}{2}-1\bigr)k,
\end{eqnarray} 
where $k$ is the curvature parameter (\ref{kdef}) for the spatial
manifold.
The conditions for equilibrium ($\dot{a}=\ddot{a}=0$ and
setting $a=1$) are
\begin{eqnarray}
  \rho &=& \frac{1}{16\pi G}n(n-1)k,  \\
  p &=& -\frac{n-2}{n}\rho. \label{equilibriump}
\end{eqnarray}
which bear little relationship to the condition $p = 0$ obtained
by extremization of total matter energy.

To determine whether the equilibrium solution is stable we examine the
linearized equations of motion.  We expand all variables to
first order: $a=1+\delta a$, $\rho=\bar{\rho}+\delta\rho$, and
$p=\bar{p}+\delta p$. Then
\begin{eqnarray}
  0 &=& 8\pi G\delta\rho +n(n-1)k\,\delta a, \\
  (n-1)\ddot{\delta a}
        &=& -8\pi G\delta p +(n-1)(n-2)k\,\delta a \nonumber \\
        &=& -8\pi G\frac{dp}{da}\delta a +(n-1)(n-2)k\,\delta a. 
        \label{equilibriumplin}
\end{eqnarray}

To obtain a stark contradiction to the non-gravitational analysis, set
$n=2$. The second of the equations for equilibrium 
(\ref{equilibriump}) then reduces to
$p=0$, which is precisely what was obtained by energy extremization.
Therefore, we may compare the general relativistic conditions for stability 
with those that would be predicted by energy minimization.
The second of the equations for stability (\ref{equilibriumplin}) gives
\begin{equation}
  \ddot{\delta a} = -8\pi G\frac{dp}{da}\delta a.
\end{equation}
In other words, this $(2\!+\!1)$-dimensional spacetime is stable only
when $dp/da$ is {\it positive\/},
exactly the opposite of eq.~(\ref{enminstab}). Thus, in this model,
the conditions for stable equilibrium in general relativity are that
the total energy of matter be a {\em maximum}.

\section{Solutions with multiple compact spaces}

In this section, we generalize the discussion of sections 2 and 3 to 
the case of multiple maximally symmetric
$n\ii$-dimensional spaces.
We begin by characterizing static solutions
as in section 2.  We then recast the full gravitational action as the
action for a set of scalar fields evolving in a four-dimensional
spacetime as in section 3, and discuss the possibility of stable
solutions.  Multiple compact spaces have been previously
considered in \cite{gz1,gz2}; to this discussion we add an
argument that the extra dimensions should be positively curved,
and we consider the height of the barrier separating the static
solution from one in which the scale factor for the extra dimensions
runs away to infinity.

Our consideration of static spacetimes is a straightforward
generalization of that in section 2, so we will be concise.
Consider a spacetime with spatial sections described by products
of $m$ maximally symmetric manifolds, such that the metric may be 
written
\be\label{eq:metric}
  ds^2 = -dt^2 + \sum_{i=1}^{m}
     \gamma^{(i)}_{\alpha_i\beta_i}dx^{\alpha_i} dx^{\beta_i},
\ee
where the $\gamma^{(i)}_{\alpha_i\beta_i}$'s are metrics of
constant curvature.  The indices
$\{\alpha_i, \beta_i\}$ run from $n_{i-1}+1$ to $n\ii$ with $n_0
= 0$, i.e. the set of indices ``runs over the submanifold ${\cal{M}}\ii$''.
The curvature parameters $k\ii$ are defined analogously to 
(\ref{kdef}).
For a stress-energy source, we consider a perfect fluid defined by an
energy density $\rho$ and a pressure $p\ii$ in each set of dimensions:
\bea
  & T_{00} = \rho \\ 
  & T_{\alpha_i\beta_i} =  p\ii \gamma^{(i)}_{\alpha_i\beta_i}.
\eea
There are $m+1$ independent components of Einstein's equation:
the $00$ component,
\be
 \label{E1}
   \sum_j\frac{1}{2}n\jj(n\jj-1)k\jj = 8\pi G\brho ,
\ee
and the $\alpha_i\beta_i$ components,
\be
 \label{E2}
  -(n\ii-1)k\ii + \sum_j\frac{1}{2}n\jj(n\jj-1)k\jj
  = -8\pi G \bp\ii ,
\ee
where once again overbars remind us that we are considering static
solutions, and $G$ is Newton's constant in the full spacetime.

The null energy condition implies
\be 
 \rho + p\ii \geq 0,
\ee
while the conditions (\ref{E1}-\ref{E2}) lead to the relation
\be
\label{krel}
  8\pi G(\brho + \bp\ii) = (n\ii-1)k\ii .
\ee
Thus, the left hand side of (\ref{krel}) is
nonnegative, and there are no static solutions with any $k\ii < 0$
that obey the NEC, in accordance with the general
argument given at the end of section 2.  None of
the spatial factors can be negatively curved; it is not possible
to compensate for some negative curvature in one factor with
additional positive curvature somewhere else.
As before, if all of the curvature parameters vanish, we recover
empty Minkowski space.  The only static nonempty solutions will
involve no negatively curved spaces and at least one positively
curved space.

With the criteria for static solutions in hand, we will now explore
the stability of these solutions to small perturbations in the scale
factors.  Following section 3, we first consider the conventional
dimensional reduction of this theory, rewriting the $4+\Sigma n\ii$
dimensional Lagrangian as a four dimensional Lagrangian plus $M$
scalar fields representing the size of each set of extra
dimensions. (The sum is only over the extra dimensions, and
$M$ refers to the number of sets of extra dimensions.)

Let $G_{ab}$ be the metric for the full spacetime with coordinates
$X^a$, and denote by $G_{4+\cal{N}}$ Newton's constant in the full
spacetime where ${\cal{N}}=\Sigma n\ii$, the number of extra
dimensions.  The indices $a$, $b$ run from 0 to $3+\cal{N}$.  We
consider metrics of the form
\be
\label{eq:gen_metricansatz}
ds^2 = G_{ab}dx^adx^b = 
	g_{AB} dx^{A} dx^{B} 
        + \sum_{i=1}^M \bi^2(x)
        \gamma^{(i)}_{\alpha_i\beta_i}(y) dy^{\alpha_i}dy^{\beta_i}
\ee
where $A$ and $B$ run over the four indices of the 3+1
dimensional spacetime and each set \{$\alpha_i,\beta_i$\} run over the
appropriate submanifold.
The action is as before:
\begin{equation}\label{eq:genHilbertAction} 
  S = \int d^{4+\cal{N}}X \sqrt{-G}\left\{\frac{1}{16\pi
	G_{4+\cal{N}}} R[G_{ab}]+\Lhat_M\right\} .
\end{equation}
From the metric ansatz
(\ref{eq:gen_metricansatz}) we have 
\bea
  \sqrt{-G} &=&  \sqrt{-g}\prod_i \bi^{n\ii}\sqrt{\gamma^{(i)}}, \\
  R[G_{ab}] &=& R[g_{AB}] + \sum_{i=1}^M \Bigl[
  \bi^{-2} R[\gamma^{(i)}] -2n\ii \bi^{-1}
  g^{AC}\nabla_A\nabla_C\bi  \\
&&\qquad  - n\ii(n\ii-1) \bi^{-2} g^{AC}(\partial_A\bi)
  (\partial_C\bi) - \sum_{j\neq i}n\ii n\jj \bi^{-1}b_j^{-1} 
	g^{AC}\partial_A\bi\partial_C b_j \Bigr] ,\nonumber
\eea
where $\nabla_A$ is the covariant derivative associated with $g_{AB}$.
The volume of the extra dimensions when all $\bi = 1$ is given by
\be
\label{volume}
  \mathcal{V} = \prod_i \int d^{n\ii}y \, \sqrt{\gamma^{(i)}} ,
\ee
and the four-dimensional Newton's constant $G_4$ is related to its
higher-dimensional version by
\be
  \frac{1}{16\pi G_4} = \frac{\mathcal{V}}{16\pi G_{4+\cal{N}}}.
\ee
We are thus left with
\bea
  S &=& \int d^{4}x \sqrt{-g}\Biggl\{\frac{\prod_i\bi^{n_i}}{16\pi G_4} 
  \Bigl[ R[g_{AB}] + \sum_{i=1}^M \bigl[
  n\ii(n\ii-1)k\ii\bi^{-2} -2n\ii \bi^{-1}
  g^{AC}\nabla_A\nabla_C\bi \nonumber \\ 
&&\qquad\qquad- n\ii(n\ii-1) \bi^{-2} g^{AC}(\partial_A\bi)
  (\partial_C\bi) - \sum_{j\neq i}n\ii n\jj \bi^{-1}b_j^{-1} 
	g^{AC}\partial_A\bi\partial_Cb_j \bigr]\Bigr] \nonumber \\
&&\qquad\qquad  + \mathcal{V} \Lhat_M\Biggr\}, 
\eea
where we have introduced the curvature parameter $k\ii$ of each set of extra
dimensions via $R[\gamma^{(i)}] = n\ii(n\ii-1)k\ii$.

Through a change of variables and a
conformal transformation, 
\bea
  \ati(x) &=& \ln{\bi(x)} - \lambda \\
  g_{ab} &=& e^{2\lambda}\tilde{g}_{ab},
\eea
where $\lambda = -\frac{1}{2}\Sigma n\ii\ln{\bi}$,  we can turn this
reduced action into that of $M$ scalar fields 
coupled to gravity in the Einstein frame.  Ignoring a total
divergence term, after some algebra the reduced action becomes
\bea \label{eq:gen_reducedaction}
S &=& \int \sqrt{-\tilde{g}}\Biggl\{
	\frac{1}{16\pi G_4}\Biggl[
	\tilde{R}[\tilde{g}_{AB}]+ \sum_i
		e^{-2\ati}n\ii(n\ii-1)k\ii \\
&& \qquad\qquad + \sum_{i,j}\left(
		 \eta_{ij}\tilde{g}^{AB}
	\tilde{\nabla}_A\ati\tilde{\nabla}_B\alpha_j\right)\Biggr]
		+ \mathcal{V}e^{-\frac{2}{2+\mathcal{N}}\Sigma n\ii\ati}
		\mathcal{L}_M\Biggr\} , \nonumber 
\eea
where
\be
  \eta_{ij} = n\ii n\jj/({2+\cal{N}}) - n \ii \delta_{ij}\ .
\ee 
In order to have terms that look like the canonical kinetic terms for
a set of scalar fields, we must diagonalize the quadratic form
$\eta_{ij}$; this has been performed explicitly by
\cite{gz1,gz2}.  Thus, we
may rewrite the action as
\bea 
S &=& \int \sqrt{-\tilde{g}}\Biggl\{
	\frac{1}{16\pi G_4}\Biggl[
	\tilde{R}[\tilde{g}_{AB}]+ \sum_i
		e^{-2\ati}n\ii(n\ii-1)k\ii \\
&& \qquad\qquad + \sum_{i}\left(
		 \eta'_{i}\tilde{g}^{AB}
		\tilde{\nabla}_A(\ati')\tilde{\nabla}_B(\ati')\right)\Biggr]
		+ \mathcal{V}e^{-\frac{2}{2+\mathcal{N}}\Sigma n'\ii\ati'}
		\mathcal{L}_M\Biggr\} \nonumber 
\eea
where $\eta'_{i}$ are the eigenvalues of $\eta_{ij}$ and the $\ati'$ are 
combinations of the $\ati$ that correspond to the $\eta'_{i}$.
Since each $\ati$ may be written as a linear combination of the
$\ati'$, we have an action for $M$ scalar fields on a background
spacetime.  The potential for these fields is positive for
$\mathcal{L}_M > 0$, although from this form it is not clear 
in general whether the static point is stable.

As an example, consider the spacetime ${\mathcal M}\times S^2 \times
S^2$, where ${\mathcal M}$ is a 3+1 dimensional homogeneous, isotropic
model \cite{gz1}. The action for this spacetime reduces to  
\bea 
S &=& \int\sqrt{-g}\bigl(\frac{1}{2}\mpl^2R[g_{AB}] 
	-{1\over 2}g^{AB}\partial_A\phi\partial_B\phi 
	-{1\over 2}g^{AB}\partial_A\psi\partial_B\psi \nonumber \\ 
&& \qquad\qquad 
     + e^{-(\sqrt{3}\phi+\psi)/\mpl}\mpl^2k_f
     +e^{-(\sqrt{3}\phi-\psi)/\mpl}\mpl^2k_h
     - {\mathcal V}e^{-\frac{2}{\sqrt{3}}\phi/\mpl}{\mathcal{L}_M}\bigr) ,
\eea 
where ${\mathcal V}$ is the volume of the extra dimensions as defined in
eq. (\ref{volume}) above, and we have introduced
\bea \phi &=& \frac{1}{\sqrt{3}}(\alpha_f+\alpha_h)\mpl \\ 
\psi &=& (\alpha_f-\alpha_h)\mpl ,
\eea 
where $f$ and $h$ label the spheres.

We choose the matter for this model to consist of a cosmological
constant, $\lambda_8$, and abelian gauge fields, $F_{ab}$ and
$H_{ab}$, so the matter Lagrangian is
\be
{\mathcal L}_M = -\lambda_8 - {1\over 4}G^{ac}G^{bd}F_{ab}F_{cd}
		   - {1\over 4}G^{ac}G^{bd}H_{ab}H_{cd},
\ee
We take the gauge fields to wrap around each $S^2$, so that,
in components, we have
\bea
F_{45} &=& -F_{54} = \sqrt{\gamma^{(f)}}F_0 \\
H_{67} &=& -H_{76} = \sqrt{\gamma^{(h)}}H_0,
\eea
where $F_0$ and $H_0$ are constants, and all other components vanish.
The conditions for a static solution are [see eqs. (\ref{E1},\ref{krel})] 
\bea
\label{static20}
8\pi G_8\bar{\rho} &=& 3K+k_f+k_h \\
\label{static21}
8\pi G_8(\bar{\rho}+\bar{P}) &=& 2K \\
\label{static22}
8\pi G_8(\bar{\rho}+\bar{p_f}) &=& k_f \\
\label{static23}
8\pi G_8(\bar{\rho}+\bar{p_h}) &=& k_h ,
\eea
where $G_8$ is the 8-dimensional Newton's constant, $K$ is the
curvature parameter for the large dimensions, and $k_f$ and $k_h$ are
the curvature parameters for $\gamma^{(f)}_{ij}$ and $\gamma^{(h)}_{ij}$,
respectively. Consequently, we find that 
static solutions exist, provided that
\bea
\label{constraint20}
\lambda_4 &=& {1\over 2}\mpl^2(k_f+k_h) \\
K &=& 0 \\
\label{constraint21}
f_0^2 &=& {1\over 2}\mpl^2k_f \\
\label{constraint22}
h_0^2 &=& {1\over 2}\mpl^2k_h ,
\eea
where
\bea
f_0^2 &=& {1\over 2}{\mathcal V}F_0^2 \\
h_0^2 &=& {1\over 2}{\mathcal V}H_0^2.
\eea
Thus, there exists a 2-parameter
family of static solutions. It is convenient to choose as
independent parameters the higher-dimensional (reduced) Planck mass 
and the ratio of the static curvature parameters of the two-spheres:
\begin{eqnarray}
  M_8^{-6} &=& \sqrt{8\pi G_8} \\
  r &=& {k_f \over k_h} .\label{r}
\end{eqnarray}
In terms of these, the equilibrium volume of the internal space is
set by
\be
{\mathcal V} = {\mpl^2 \over M_8^6} ,
\ee
where $\mpl = 10^{18}$~GeV is the (reduced)
four-dimensional Planck scale.  We will assume
that $M_8 =  1$~TeV; this implies ${\mathcal V} \approx (30~{\rm keV})^{-4}$.
The curvature parameters for the spheres can be determined from
(\ref{r}) and
\be
  k_f k_h = {16\pi^2 \over {\mathcal V}}\ ;
\ee
we have for example
\be
  k_f = 4\pi {M_8^3 \over \mpl} \sqrt{r}. \label{kf}
\ee

The potential in this matter model is  
\begin{eqnarray}
V(\phi,\psi) &=& 2\pi\mpl M_8^3
   \Biggl[\left({1+r \over \sqrt{r}}\right)e^{-(2/\rt\mpl)\phi} 
   + \sqrt{r}\left(e^{-2[(2/\rt)\phi+\psi]/\mpl}
		-2e^{-(\rt\phi+\psi)/\mpl}\right) \nonumber \\
   &&\qquad\qquad\qquad\qquad
    + {1\over \sqrt{r}}\left(e^{-2[(2/\rt)\phi-\psi]/\mpl}
		-2e^{-(\rt\phi-\psi)/\mpl}\right)\Biggr].
		\label{multipot}
\end{eqnarray}
The fields which diagonalize the mass matrix 
at the equilibrium solution ($\phi=0$, $\psi =0$) are 
\begin{eqnarray}
\Phi &=& \sqrt{\frac{2-\mphi}{\mpsi-\mphi}}\left(
	\phi-\frac{2(r-1)}{\rt(r+1)(2-\mphi)}\psi\right) \\
   &\,& \nonumber \\
\Psi &=& \sqrt{\frac{\mpsi - 2/3}{\mpsi-\mphi}}\left(
	\frac{2(r-1)}{\rt(r+1)(\mpsi-2/3)}\phi+\psi\right),
\end{eqnarray}
where the $\mu$'s are normalized masses,
\begin{eqnarray}
  \mphi &=& {m_\Phi^2\mpl^2 \over \lambda_4} =
	\frac{4}{3}\left(1-\sqrt{1-\frac{3r}{(1+r)^2}}\right)
    \\
  \mpsi &=& {m_\Psi^2\mpl^2\over \lambda_4} = 
	\frac{4}{3}\left(1-\sqrt{1+\frac{3r}{(1+r)^2}}\right)
   \ ,
\end{eqnarray}
and $m_\Phi$ and $m_\Psi$ are the actual masses of the $\Phi$ and $\Psi$
fields respectively.

It is clear that replacing $r \rightarrow 1/r$ is equivalent to
interchanging the roles of the $F$ and $H$ gauge fields; we therefore
only consider $r\geq 1$, or $k_f \geq k_h$ (so that the sphere with
$F$-flux is {\em smaller} than that with $H$-flux).
There is also an upper limit on $r$, derived
from the requirement that the curvature of the extra dimensions
not exceed the higher-dimensional Planck scale, $k_f \leq M_8^2$.
From (\ref{kf}), setting $M_8 = 1$~TeV then implies
\be
  r \leq 10^{28} .
\ee
There is a complementary constraint, that the larger two-sphere not
be big enough that deviations from Newtonian gravity would have been
detected in laboratory experiments; this corresponds to 
$k_h < (10^{-2}~{\rm eV})^{-2}$ \cite{exp}.  This bound is just 
saturated when $r = 10^{28}$, so these bounds are 
essentially equivalent.

For finite values of $r$, the masses of both $\Phi$ and $\Psi$
are positive, implying stability of the system around the equilibrium
solution.  We would like to comment briefly on the cosmological
acceptability of such a scenario, following the discussion of the
height of the potential barrier investigated in section~\ref{example}.
For a given value of $r$, the relevant height is that of the
lowest barrier separating the equilibrium solution from infinity.
This will correspond to a saddle point in $V(\Phi, \Psi)$, and the
barrier height will be the value of $V(\Phi,\Psi)$ evaluated
at the saddle point.
Numerically investigating the potential (\ref{multipot}) with 
$M_8 = 1$~TeV, we find that 
the height of the barrier varies with $r$, but  
the barrier always remains at least 1~TeV$^4$.  Since
this is the fundamental scale of quantum gravity, we again find that
there is no Brustein-Steinhardt problem \cite{brusteinsteinhardt}
for this theory.  At energy scales where classical gravity should
be applicable, the field will never be able to classically
roll over the barrier. 

\section*{Acknowledgments}
We thank Mark Trodden for helpful conversations.  This work was
supported in part by the NSF grants PHY 95-14726, PHY 01-14422,
and PHY 00-90138, and DOE grant DE-FG02-90ER-40560 to
the University of Chicago,  the Alfred
P. Sloan Foundation, and the David and Lucile Packard Foundation.
One of us (R.M.W.) would like to
thank the Yukawa Institute of Kyoto University for its hospitality.

\end{document}